\documentclass{appolb}
\usepackage{graphicx}
\usepackage[T1]{fontenc}
\usepackage[latin9]{inputenc}
\usepackage{cite}
\usepackage{amsmath}
\usepackage{amssymb}

\newcommand{\CD}{{\cal D}}

\newcommand{\CI}{{\cal I}}

\newcommand{\CR}{{\cal R}}

\newcommand{\CQ}{{\cal Q}}

\newcommand{\average}[1]{\left\langle #1 \right\rangle_\CD}

\newcommand{\caverage}[1]{\left\langle #1 \right\rangle_{\CI}}

\newcommand{\initial}[1]{{{#1}_{\mathbf i}}}

\newcommand{\inI}{{\mathrm{I}}}
\newcommand{\inII}{{\mathrm{II}}}
\newcommand{\inIII}{{\mathrm{III}}}

\providecommand{\eprint}[1]{\href{http://arxiv.org/abs/#1}{{\tt [arXiv:#1]}}}

\usepackage{hyperref}

\begin{document}
\title{Mass Function of Galaxy Clusters\\ in Relativistic Inhomogeneous Cosmology%
\thanks{Presented at the 3rd Conference of the Polish Society on Relativity}%
}
\author{Jan J. Ostrowski$^1$, Thomas Buchert$^1$, Boudewijn F. Roukema$^2$
\address{$^1$Univ Lyon, Ens de Lyon, Univ Lyon1, \\CNRS, Centre de Recherche Astrophysique de
Lyon UMR5574, F--69007, Lyon, France}
\address{$^2$ Toru\'n Centre for Astronomy,
  Faculty of Physics, Astronomy and Informatics,
  Grudziadzka 5,
  Nicolaus Copernicus University, ul. Gagarina 11, 87-100 Toru\'n, Poland}
}
\headtitle{Mass Function of Galaxy Clusters in Relativistic Inhomogeneous Cosmology}
\headauthor{Ostrowski, Buchert \& Roukema}

\maketitle
\begin{abstract}
The current cosmological model ($\Lambda$CDM) with the underlying FLRW metric relies on the assumption of local isotropy, hence homogeneity of the Universe. Difficulties arise when one attempts to justify this model as an average description of the Universe from first principles of general relativity, since in general, the Einstein tensor built from the averaged metric is not equal to the averaged stress--energy tensor. In this context, the discrepancy between these quantities is called ``cosmological backreaction'' and has been the subject of scientific debate among cosmologists and relativists for more than $20$ years. Here we present one of the methods to tackle this problem, i.e. averaging the scalar parts of the Einstein equations, together with its application, the cosmological mass function of galaxy clusters.
\end{abstract}

\section{Introduction}
The cosmological backreaction problem is usually discussed in the context of large scales (the homogeneity scale), by the means of (i) perturbation theory, (ii) exact solutions of the Einstein equations, together with averaging methods. Averaging methods are especially useful on smaller scales where the perturbative approach breaks down. Here, we propose an application of scalar averaging to dark matter (DM) halo formation, and in particular, we derive the semi-analytic DM halo mass function. Our method provides, in principle, an observational test of the scalar averaging approach.

\section{Scalar-averaged Einstein equations and the mass function}
One way to capture the discrepancies between the evolution of the averaged and homogeneous variables is to average the scalar parts of the Einstein equations for irrotational dust in a $3+1$ comoving-synchronous foliation, leading to the effective acceleration and expansion laws, e.g. \cite{Bgrg},
\begin{equation}
\label{ray}
3\frac{\ddot{a}_{\CD}}{a_{\CD}} + 4\pi G \frac{M_{\CD_i}}{V_{\CD_i}a^3_{\CD}} = \mathcal{Q}_{\CD} \;, \\
\end{equation}
\begin{equation}
\label{ham}
\left(\frac{\dot{a}_{\CD}}{a_{\CD}}\right)^2 - \frac{8\pi G}{3}\frac{M_{\CD_i}}{V_{\CD_i}a^3_{\CD}} + \frac{\langle \mathcal{\CR}\rangle_{\CD}}{6} = -\frac{\mathcal{Q}_{\CD}}{6}\;,
\end{equation}
subject to the integrability condition
\begin{equation}
\frac{1}{a_\CD^6}(\,{\CQ}_\CD \,a_\CD^6 \,)\,\dot{}
\;+\; \frac{1}{a_\CD^{2}} \; (\,\average{\CR}a_\CD^2 \,)\,\dot{}\,\;=\;0\;,
\label{int}
\end{equation}
where $a_{\CD}$ is the effective scale factor, defined through the volume $V_\CD$ of the domain of averaging,
\begin{equation}
a_{\CD} = \left(\frac{V_\CD (t)}{V_{\CD_i}}\right)^{\frac{1}{3}},
\end{equation}
and $M_{\CD_i}$ stands for the initial mass of the domain (conserved by assumption during the evolution). Terms describing the effects of inhomogeneities are the averaged spatial scalar curvature $\average{\CR}$ and the kinematical backreaction:
\begin{eqnarray}
  \CQ_{\CD} &=& \frac{2}{3}\left(\average{\Theta^2} - \average{\Theta}^2\right) - 2\average{\sigma^2}
  \,=\, 
  2\average{{\rm II}(K^i_{\; j})} - \frac{2}{3}\average{{\rm I}(K^i_{\; j})}^2,
\end{eqnarray}
where in the first line we use two scalar variables,
the rate of expansion $\Theta$ and the rate of shear $\sigma$,
from the kinematical decomposition of the extrinsic curvature tensor (in mixed index notation):
\begin{equation}
K^i_{\;j} := \frac{1}{2}g^{ik}\dot{g}_{kj}\;\;;\;\;\Theta:= -K^k_{\;k} \;\;;\;\;\sigma^i_{\;j} = -K^i_{\;j} - \frac{1}{3}\Theta\delta^i_{\;j}\;\;;\;\;\sigma := \frac{1}{2} \sigma^i_{\; j}\sigma^j_{\; i}\;,
\end{equation}
where the second line features two of the principal scalar invariants of the extrinsic curvature matrix.

\subsection{Relativistic Zel'dovich approximation (RZA)}
The relativistic version of the Zel'dovich approximation (RZA) follows and extrapolates the logic of original, classical work (see \cite{RZA1} for its definition and references). The RZA is based on the $3+1$ splitting of the Einstein equations within a comoving-synchronous foliation. In analogy to the Newtonian Lagrangian perturbation theory, the only variable in this set of equations is the relativistic equivalent of the classical fluid deformation gradient, represented by Cartan co-frame fields, which is decomposed into the homogeneous isotropic background and the inhomogeneous part. By inserting a restricted sub-case of the first-order solution of the Einstein equations into the functionals for averaged fields without any further linearization, we  obtain non-perturbative expressions in the spirit of the original Zel'dovich expression for the density.
For instance, the backreaction term in the RZA takes the following form \cite{RZA2}:
\begin{eqnarray}
 {}^{{\rm RZA}}\CQ_{\CD}\;=  {\displaystyle
    \frac{\dot{\xi}^{2}\left(\gamma_{1}+\xi\gamma_{2}+\xi^{2}\gamma_{3}\right)}{\left(1+\xi\caverage{{\rm
          I}_{{\rm {\bf i}}}}+\xi^{2}\caverage{{\rm
          II}_{{\rm {\bf i}}}}+\xi^{3}\caverage{{\rm
          III}_{{\rm {\bf i}}}}\right)^{2}}\;,} \quad {\rm with:} \qquad\nonumber\\
  \begin{cases}
  &  \gamma_{1}:=2\caverage{{\rm II}_{{\rm {\bf i}}}}-\frac{2}{3}\caverage{{\rm I}_{{\rm {\bf i}}}}^{2}\\
  &  \gamma_{2}:=6\caverage{{\rm III}_{{\rm {\bf i}}}}-\frac{2}{3}\caverage{{\rm II}_{{\rm {\bf i}}}}\caverage{{\rm I}_{{\rm {\bf i}}}}\\
   & \gamma_{3}:=2\caverage{{\rm I}_{{\rm {\bf i}}}}\caverage{{\rm III}_{{\rm {\bf i}}}}-\frac{2}{3}\caverage{{\rm II}_{{\rm {\bf
          i}}}}^{2}\;,
\end{cases}
\label{QRZA}
\end{eqnarray}
where
$\caverage{\initial{\inI}}$, $\caverage{\initial{\inII}}$ and $\caverage{\initial{\inIII}}$ are the initial values of the principal scalar invariants introduced above, averaged over the initial domain; $H$ is the background Hubble expansion rate, and $\xi$ is a time-dependent function related to the assumed background. Thus, the RZA provides the closure condition for the set of scalar averaged equations $\lbrace (\ref{ray}), (\ref{ham}), (\ref{int}) \rbrace$, and reduces the problem to equations for the effective scale factor and the initial value problem.

\subsection{Mass function of dark matter halos}
\label{massfunction}
The mass function (MF) of DM is the number density of collapsed objects per mass interval at a given redshift (see e.g. \cite{Monaco} for a review and references to the original work on this subject).
In practice, we usually use the differential number density of halos per logarithmic halo mass:
\begin{eqnarray}
\frac{\mathrm{dN}}{\mathrm{d} \ln M} = \frac{\varrho_0}{M}\;f(\sigma_M) \left|\frac{\mathrm{d} \ln \sigma_M}{\mathrm{d} \ln M}\right| \;,
\end{eqnarray}
where $\varrho_0$ is the background density, $\sigma_M$ is the mass variance and $f(\sigma_M)$ is called the multiplicity function. To calculate $f(\sigma_M)$ in our setup we first calculate the distributions of initial averaged invariants (see \cite{BKS}) on initially spherical domains. Next, we generate $2^{15}$ realizations of these distributions for each mass scale, and numerically solve the averaged Raychaudhuri equations (with the RZA ansatz) for each of these cases. We then focus on the fraction of collapsed objects (fulfilling the condition $a_{\CD} \approx 0$) at each mass scale, thus obtaining the cumulative distribution function of DM halos. We nonlinear--least-squares fit this with an exponential and differentiate with respect to mass. We thus obtain the probability density function from which the multiplicity function is calculated.

\section{Results and conclusion}
To check the influence of small inhomogeneities on the average expansion, we also
performed the subsection \ref{massfunction} calculations
for the case of spherical symmetry, obeying the conditions $\gamma_1 = \gamma_2 = \gamma_3 = 0$, see
(\ref{QRZA}), which is an exact sub-case of RZA, and also for the restricted case given by
$\caverage{{\rm II}_{{\rm {\bf i}}}} = \caverage{{\rm III}_{{\rm {\bf i}}}}= 0$.
\begin{figure}[htb]
\centerline{%
\includegraphics[width=0.5\textwidth]{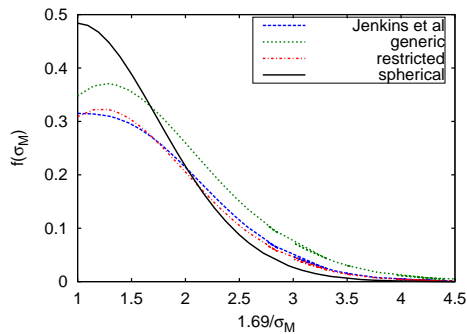}}
\caption{Multiplicity functions for Jenkins et al.'s \cite{Jenkins} $N$-body fit (dashed line), generic and restricted initial invariants (dotted and dashed dotted lines respectively), and spherical collapse (solid line) at $z=0$.}
\label{Fig:F2H}
\end{figure}
Figure~\ref{Fig:F2H} shows the close resemblance between the restricted case and the Jenkins et al. \cite{Jenkins} fitting function. It also shows a big discrepancy between the spherical case and the other
three cases.
Neither the sphericity assumption nor the approach of \cite{Jenkins} match the more realistic generic
case\footnote{{\it Acknowledgments:} The work of TB and JJO was conducted within  the ``Lyon Institute of
Origins'' under grant  ANR-10-LABX-66. Part of this work was funded by the National Science Centre, Poland, under grant 2014/13/B/ST9/00845. Computations were made under grant 197 of the
Pozna\'n Supercomputing and Networking Center.}. 

\end{document}